\documentstyle{article}
\begin{document}

\noindent
{\Large Erratum to ``A phenomenological description of
$\pi^-\Delta^{++}$ photo- and electroproduction in nucleon resonance
region''
[Nucl. Phys. A672 (2000) 220-248]}

\vspace*{1pc}
\noindent
M. Ripani $^a$, V. Mokeev $^b$, M. Anghinolfi $^a$, M. Battaglieri $^a$,
G. Fedotov $^c$, E. Golovach $^{a,b}$, B. Ishkhanov $^{b,c}$,
M. Osipenko $^c$, G. Ricco $^{a,d}$, V. Sapunenko $^{a,b}$ and
M. Taiuti $^{a,d}$

\vspace*{1pc} \noindent
{\it $^a$ Istituto Nazionale di Fisica
Nucleare, Via Dodecanneso 33,
     I-16146 Genova, Italy\\
$^b$ Nuclear Physics Institute, Moscow State University,
     Vorob'evy gory, 119899 Moscow, Russia\\
$^c$ Physical Faculty of Moscow State University, Vorob'evy gory,
     119899 Moscow, Russia\\
$^d$ Dipartimento di Fisica, Universit\`a di Genova, Via Dodecanneso 33,
     I-16146, Genova, Italy\\}

\vspace{1pc}

When our paper \cite{Rip} was about to be printed,
we learned that a similar work was going to be
published by J.C. Nacher, E. Oset (University of
Valencia): this paper actually appeared as \cite{Nac}.
We would like to acknowledge this paper as an important
contribution to the subject, being the model described
very accurate in the description of $\Delta \pi$
electroproduction in the  low invariant mass region,
i.e. $W$ below 1.6 GeV.
In particular, this paper contains an interesting
discussion on gauge invariance and on the effect of
the choice of form factors on the observables; also,
interference effects between the resonant state
$D_{13}(1520$ and the continuum are discussed in detail.
We would like also to point out that our discussion
of unitarity issues in \cite{Rip} was meant to
underline that other models preceding ours did not
contain an explicit treatment of the coupling to
competitive channels through the use of hadronic
amplitudes appropriate for the resonance region.
On the other hand, in the work preceding \cite{Nac}
and also reported in the bibliography of \cite{Rip},
some discussion of unitarity was included, showing
that the approach developed in those papers is not
expected to violate unitarity significantly
in the region below 1.6 GeV invariant energy.

\vfill

\end{document}